\documentclass[12pt]{article}

\usepackage{hyperref}       
\usepackage{url}            
\usepackage{booktabs}       
\usepackage{amsfonts, amsmath, amsthm, amssymb, dsfont,mathrsfs,mathabx,amsbsy}
\usepackage{nicefrac}       
\usepackage{microtype}      
\usepackage{bm}
\usepackage{pdfpages}
\usepackage{enumerate}
\usepackage{float}
\usepackage{natbib}
\usepackage{fullpage}
\usepackage{caption}
\usepackage{setspace}
\usepackage{tabularx}
\newcommand\setrow[1]{\gdef\rowmac{#1}#1\ignorespaces}
\newcommand\clearrow{\global\let\rowmac\relax}
\clearrow

\newcommand{\N}{\mathbb{N}}

\newcommand{\Pb}{\operatorname{P}}
\newcommand{\iid}{\overset{iid}{\sim}}
\newcommand{\ind}{\overset{ind}{\sim}}
\newcommand{\h}{\hspace{4pt}}
\newcommand{\mK}{\mathcal{K}}

\newcommand{\E}{\mathrm{E}}

\def\app#1#2{%
  \mathrel{%
    \setbox0=\hbox{$#1\sim$}%
    \setbox2=\hbox{%
      \rlap{\hbox{$#1\propto$}}%
      \lower1.3\ht0\box0%
    }%
    \raise0.25\ht2\box2%
  }%
}
\def\approxprop{\mathpalette\app\relax}

\newcommand\indep{\protect\mathpalette{\protect\independenT}{\perp}}
\def\independenT#1#2{\mathrel{\rlap{$#1#2$}\mkern2mu{#1#2}}}

\theoremstyle{plain}

\theoremstyle{definition}

\theoremstyle{plain}

\theoremstyle{plain}

\newcommand{\beginsupplement}{%
        \setcounter{section}{0}
        \renewcommand{\thesection}{S\arabic{section}}%
        \setcounter{table}{0}
        \renewcommand{\thetable}{S\arabic{table}}%
        \setcounter{figure}{0}
        \renewcommand{\thefigure}{S\arabic{figure}}%
        \setcounter{definition}{0}
        \renewcommand{\thedefinition}{S\arabic{section}.\arabic{definition}}%
        \setcounter{lemma}{0}
        \renewcommand{\thelemma}{S\arabic{section}.\arabic{lemma}}%
        \setcounter{theorem}{0}
        \renewcommand{\thetheorem}{S\arabic{section}.\arabic{theorem}}%
     }

\title{Coarsened mixtures of hierarchical skew normal kernels for flow cytometry analyses}

\author{
  Shai Gorsky\\
  Duke University\\
  Durham, NC 27708\\
  \texttt{s.gorsky@duke.edu} \\
  \and
  Cliburn Chan\\
  Duke University\\
  Durham, NC 27708\\
  \texttt{cliburn.chan@duke.edu}
  \and
  Li Ma \\
  Duke University\\
  Durham, NC 27708\\
  \texttt{li.ma@duke.edu}
}
\date{}

\begin{document}

\maketitle

\begin{abstract}
Flow cytometry (FCM) is the standard multi-parameter assay  for measuring single cell phenotype and functionality. It is commonly used for quantifying the relative frequencies of cell subsets in blood and disaggregated tissues. A typical analysis of FCM data involves cell classification---that is, the identification of cell subgroups in the sample---and comparisons of the cell subgroups across samples or conditions. While modern experiments often necessitate the collection and processing of samples in multiple batches, analysis of FCM data across batches is challenging because 
differences across samples may occur due to either true biological variation or technical reasons such as antibody lot effects or instrument optics across batches. Thus a critical step in comparative analyses of multi-sample FCM data---yet missing in existing automated methods for analyzing such data---is cross-sample calibration, whose goal is to align corresponding cell subsets across multiple samples in the presence of technical variations, so that biological variations can be meaningfully compared. We introduce a Bayesian nonparametric hierarchical modeling approach 
for accomplishing both calibration and cell classification simultaneously in a unified probabilistic manner. Three important features of our method make it particularly effective for analyzing multi-sample FCM data: a nonparametric mixture avoids prespecifying the number of cell clusters; a hierarchical skew normal kernel that allows flexibility in the shapes of the cell subsets and cross-sample variation in their locations; and finally the ``coarsening'' strategy makes inference robust to departures from the model such as heavy-tailness not captured by the skew normal kernels. 
We demonstrate the merits of our approach in simulated examples and carry out a case study in the analysis of two multi-sample FCM data sets.
\end{abstract}

\newpage

\doublespacing

\section{Introduction}
\label{sec:intro}
Flow cytometry (FCM) is a standard biological assay for measuring single cell features, referred to as ``markers'', and is commonly used for quantifying the relative frequencies of cell subsets in blood and disaggregated tissues. In this assay, individual cells labeled with fluorescent monoclonal antibodies flow single-file in a stream where they are excited by light from lasers. Light passed through (forward scatter or FSC), deflected (side scatter or SSC), as well as the characteristic wavelengths emitted by the excited fluorescent monoclonal antibodies bound to specific cell macro-molecules (typically cell surface proteins) are captured by optical detectors and reported as continuous intensity values for each marker on a cell. The general markers FSC and SSC reflect the size and internal complexity (granularity) of a cell. The fluorescent monoclonal antibodies are chosen to detect specific biological macro-molecules that are characteristic of a particular cell type. For example,a high signal from antibodies to the proteins designated as CD3 and CD4 indicate that the cell is a helper T cell while a high signal from antibodies to CD19 indicate that the cell is a B cell. Hence analysis of the relative intensity of signals from different markers is used to characterize the basic type (e.g. T or B cell), maturation (e.g. naive or memory), and activation status (e.g. quiescent or activated). Flow cytometry can scan millions of cells in solution in minutes, and provide information on the distribution of cell types in a sample, most typically reporting on the distribution of immune cell subsets in a blood sample. This information has multiple applications in clinical research, including determining the immune response to infection or vaccine challenge.

Flow cytometry suffers from  spectral overlap between light wavelengths emitted by the fluorescent dyes that limit the number of different markers that can be resolved - currently the upper limit is about 30 markers. A more recent technology based on time-of-flight spectroscopy (mass cytometry) uses heavy metal isotopes instead of fluorescent dyes to label probes thereby allowing a larger number of markers (50-100) to be measured. Because the data generated from these two technologies are very similar in nature aside from the number of markers measured simultaneously,  
herein we shall simply refer to them both as FCM data.

Traditionally, FCM data is analyzed manually by visual demarcation of cell subsets on a sequence of 2D projections, a process known as gating. Manual gating becomes unwieldy as the data dimensionality grows, and automated methods for cell subset identification from FCM data are becoming increasingly necessary, especially since the arrival of mass cytometry. However, variations in cell subset locations (in the marker space) frequently occur due to uncontrolled technical reasons unrelated to the underlying biological differences. In particular, technical differences often make it extremely challenging to compare samples processed in different batches and/or laboratories. Within a single cytometry laboratory, batch differences may occur because of instrument and reagent variability over time. These batch differences are compounded when multiple laboratories are involved in processing the data, for example, in large multi-center vaccine trials, with additional variability introduced by center-specific instruments and sample processing protocols. Thus a prerequisite for proper analysis of such data is cross-sample calibration---aligning cell subsets across multiple samples---so that cell subset properties can be meaningfully compared. But none of the existing automated classification methods performs calibration in-house and therefore cannot be directly applied to multi-sample studies. 

While we present a first approach to achieving automated classification and calibration simultaneously in a unified framework, it is worth noting that considerations of cross-sample variability, which is a key step toward calibration, have appeared in a number of previous works in the applied Bayesian nonparametrics literature for FCM data. For example, \cite{cron2013} developed an extension to the Dirichlet Process Mixture (DPM) of Gaussian kernels, in which each sample has its own set of weights, thus accounting for sample variability in subset sizes. \cite{dundar2014} further expanded the approach to also model cross-sample variability in kernel parameters by adding a DPM prior on the means of the Gaussian kernels. \cite{Lee2015} proposed a hierarchical structure where each sample is a finite mixture model of multivariate skew-$t$ distributions with random effects on the location parameters. \cite{soriano2019ba} suggested using a DPM of hierarchical Gaussian kernels to allow for variability in cell subset locations across samples. 

Building upon these works, we adopt the Bayesian nonparametric mixture framework and design a model that incorporates all of the aforementioned features of cross-sample variation. In particular, we explicitly model sample variability in relative frequencies by allowing each sample its own set of weights and incorporate a hierarchical multivariate skew normal kernels to characterize both the flexible shapes of the cell subsets as well as their variation across samples. This hierarchical kernel gives rise to a natural scheme for performing calibration, as explained in Section \ref{sec:calib}. Aside from efforts in model design, in carrying out posterior inference we incorporate a recently introduced model-robustification strategy called ``coarsening'' \citep{miller2018}. We have found that robustifying generative models for FCM data is necessitated by the complexity of such data in both the kernel features---corresponding to the shapes and heavy-tailness of cell subsets---and in the cluster weights. 
With this strategy, our method becomes robust to limitations of the skewed Gaussian kernels and mitigates the undesirable issue of DPMs in producing a diverging number of clusters as sample size grows, thereby increasing the efficacy of our model in classification and calibration. A final contribution of our framework is computational. When fitting mixtures with multivariate skew kernels, frequentist estimation with EM or Bayesian approaches with conjugate priors and Gibbs sampling face difficulties in overcoming the multimodality of the likelihood function of those kernels. We introduce a hybrid sampler that embeds a Population Monte Carlo (PMC) step into a Gibbs sampler, which reduces the risk of the sampler to be caught in a local mode and allows us to choose and sample from more reasonable though non-conjugate priors effectively.

Before describing our method in detail, we provide a quick review of some other previous statistical approaches to analyzing FCM data as well as some related literature in Bayesian modeling beyond the scope of Bayesian nonparametrics.
\cite{murphy1985} applied K-means cluster analysis to FCM data. 
Generally, K-means clustering can be very dependent on the initialization and centroid calculations, and lacks statistical interpretation. \cite{bakker1993} performed cluster analysis using K-means initialized with a large number of seed points, followed by a modified nearest neighbor technique to reduce the large number of subclusters. This method caters to symmetric clusters. \cite{boedig2008} applied finite Gaussian mixtures to FCM data, and adopted a frequentist estimation strategy based on the expectation-maximization (EM) algorithm. \cite{chan2008} fit FCM data with a finite mixture of multivariate Gaussians using standard conjugate Bayesian analysis and Gibbs sampling for inference. Methods that model FCM data with vanilla Gaussian kernels suffer from the obvious weakness that cell clusters are typically asymmetric. \cite{malsiner2017} offered a Bayesian model that allows a finite mixture of mixtures of Gaussian kernels and demonstrate this approach on FCM data. This method automates the selection of the number of clusters, and allows asymmetric clusters. 
In the frequentist literature, \cite{pyne2010} formulated the \texttt{FLAME} framework that models FCM data with a finite mixture model of skew-$t$ distributions. \cite{ohagan2016} suggested using the multivariate normal inverse Gaussian distribution kernels in the context of finite mixture models. \cite{lo2008} proposed a finite mixture of $t$ distributions with a Box-Cox transformation in order to reduce asymmetry. This method is data- and variable-dependent, which makes full automation of the process difficult. \cite{arellano2009} discussed Bayesian mixtures of multivariate skew normal kernels, but did not provide a unified approach that handles all three parameters of the skew normal distribution. \cite{sylvia2010} developed a Bayesian, fully conjugate multivariate finite mixture model with multivariate skew normal and skew-$t$ distributions. As is further discussed in Section \ref{sec:implementation} this formulation is constrained by the strong conjugate prior structure and does not allow for intuitive treatment of calibration for FCM data. \cite{hejblum2017} used the Bayesian formulation of \cite{sylvia2010} that also accommodates dependencies within the data using a sequential sampler.

The rest of the paper is organized as follows. Section~\ref{sec:model} provides the details of our Bayesian hierarchical model, the coarsening strategy, as well as the inference recipe for achieving classification and calibration jointly. 
In Section~\ref{sec:nmrcl} we provide numerical examples on simulated data that demonstrate the efficacy of our model. 
In Section~\ref{sec:casestudies} we carry out a case study on a 6-dimensional FCM data set as well as a 19-dimensional mass cytometry data set.

\section{Method\label{sec:model}}
\subsection{A Bayesian nonparametric hierarchical model}
The complexity of the observed structures in FCM data requires flexible statistical models to characterize their key features. Our model captures those by a DPM \citep{Ferguson1983} with several modeling choices that allow sample variability in weights and kernel locations as well as flexibility in the kernel shapes. We choose to work with DPMs as they are the most simple nonparamtric mixtures, which do not limit the number of clusters \textit{a priori}. Since FCM data rarely involves symmetric clusters we choose to work with skew normal (SN) kernels. These are characterized by a location parameter $\bm\xi$, a scale parameter $\bm\Sigma$, and a skew parameter $\bm\alpha$. (See Supplementary Materials~\ref{sec:msn} for further details about the multivariate SN distribution.)

\begin{figure}[!h]
\centering
\includegraphics[width=0.4\textwidth]{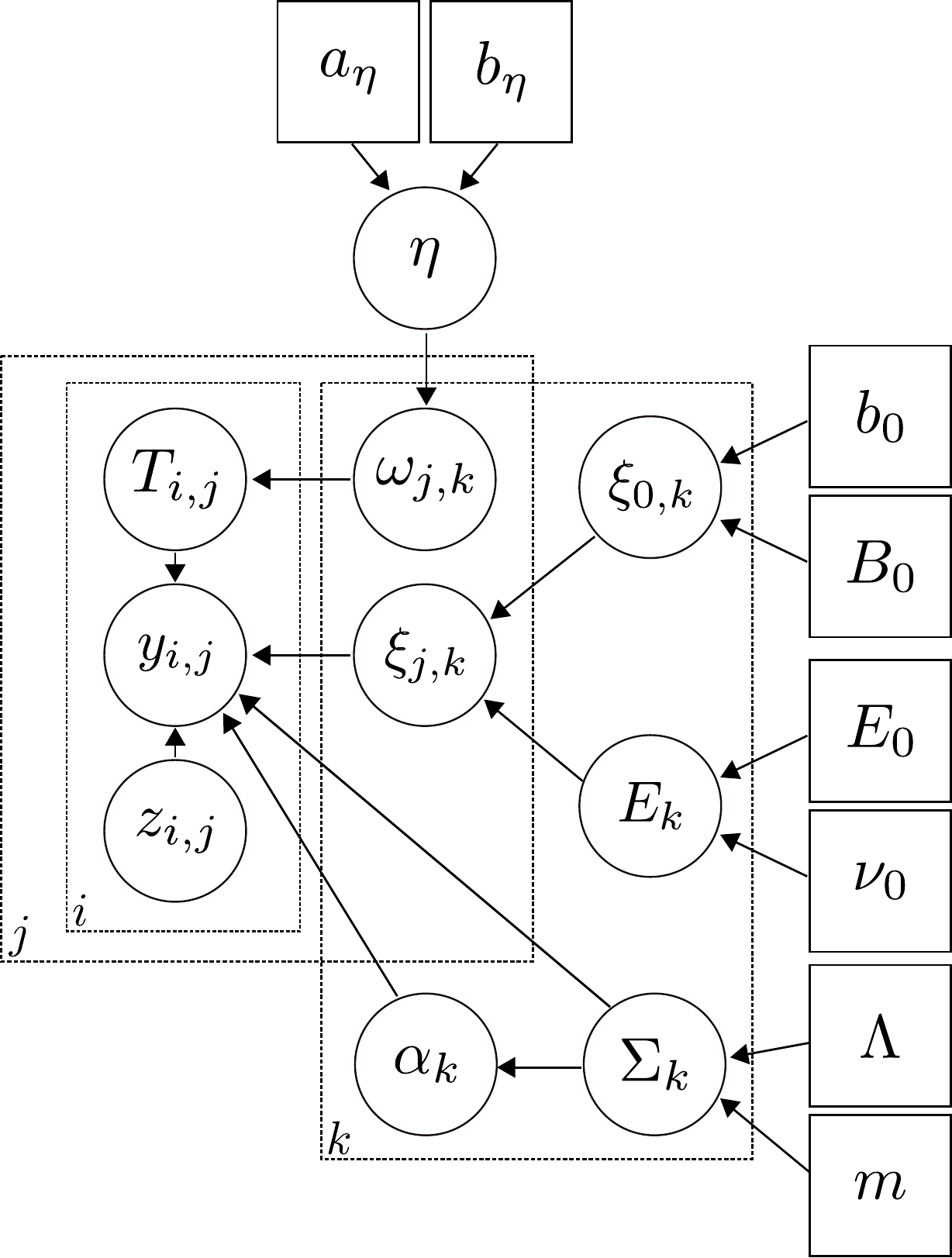}
\caption{A graphical representation of our hierarchical model. $j$ indicates the sample number; $i$ the observation number within a sample $j$; $k$ indexes cluster labels. $\bm y_{i,j}$ is the $i$th observation in sample $j$; $T_{i,j}$ the cluster assignment random variable specifying which cluster $\bm y_{i,j}$ belongs to. Each $z_{i,j}$ is associated with $\bm y_{i,j}$ and is used in augmenting the likelihood. $\bm\xi_{j,k}$ is the sample and cluster specific location parameter. $\omega_{j,k}$s are cluster and sample specific DPM weights. $\bm\alpha_k$ and $\Sigma_k$ are, respectively, the skewness vector and covariance matrix for each cluster, they are shared among all samples. $\bm\xi_{0,k}$ and $E_k$ are, respectively, the mean and the covariance of the mutlivariate normal distribution from which the sample specific means are drawn.  \label{fig:MPSK}}
\end{figure}

To allow the clusters to vary in sizes across samples, we endow each sample with its own set of mixing weights. In order to allow differences in the location of the clusters, we posit a hierarchical structure for the SN kernels. Figure~\ref{fig:MPSK} presents a full graphical view of our model. We next describe the model components in detail.

 In the following, we assume that the data are drawn from $J$ samples:
\[\bm y_{i,j}\ind F_j, \h i=1,\hdots, n_j \text{ and } j=1,\hdots, J\]
such that $n_j$ is the number of observations in sample $j\in\{1,\hdots,J\}$ and $n=\sum_{j=1}^J n_j$ is the total number of observations across all samples. 

{\em Cluster assignment and weights.}
Let $\mK$ be a countable set of cluster labels shared over all samples. For each $i=1,\hdots, n_j$ and $j=1,\hdots, J$, let $T_{i,j}$ be the latent cluster assignment variable for observation $i$ in sample $j$, such that $T_{i,j}=k$ if and only if $\bm y_{i,j}$ belongs to cluster $k$ for each $k\in\mK$. As discussed above, we allow each sample to have its own set of weights (i.e., cluster sizes), $\omega_{j,k}:=\Pb(T_{i,j}=k)$. The subscript $j$ indicates that cluster sizes may vary across samples. Note that this also allows a cluster to appear in some samples but not in others; the latter samples will simply have very small weights for that cluster. 
To form a DPM, we assign a  GEM$(\eta)$ prior \citep{Ewens1990} on the sample specific weights. For each $j\in\{1,\hdots,J\}$, let \[\{\omega_{j,k}\}_{k\in\mK}\iid \mathrm{GEM}(\eta).\]
In order to learn $\eta$, we also assign it a hyperprior, $\eta\sim {\rm Gamma}(a_\eta,b_\eta)$. 
\vspace{0.5em}

{\em Hierarchical multivariate skew normal kernels.}
Assume now that each $F_j$ has the density
\[f_j(\cdot)=\sum_{k\in\N}\omega_{j,k}\cdot g(\cdot\mid\bm\lambda_{j,k}) \]
where $g$ is the density function of the multivariate SN distribution, and each $\bm\lambda_{j,k}$ is comprised of the three parameters of the SN distribution:  $\bm\lambda_{j,k}=\{\bm\xi_{j,k}, \bm\Sigma_k,\bm\alpha_k\}$. In particular, for each cluster $k$ we assume that the scale parameter $\bm\Sigma_k$ and the skew parameter $\bm\alpha_k$ are the same across samples. We allow sample variability in the sample-specific location parameter $\bm\xi_{j,k}$ by assuming that they are distributed around a grand ``centroid'' cluster mean $\bm\xi_{0,k}$ following a multivariate Gaussian with covariance $\bm E_k$. Thus we have the following hierarchical kernel for generating an observation in each cluster
\begin{align*}
[\bm\xi_{j,k}\mid\bm\xi_{0,k}, \bm E_k, S_k ]&\ind {\rm N}(\bm\xi_{0,k}, \bm E_k)\\
[\bm y_{i,j}\mid T_{i,j}=k, \bm\xi_{j,k}, \bm\Sigma_k, \bm\alpha_k]&\sim {\rm SN}_p(\bm\xi_{j,k},\bm \Sigma_k, \bm\alpha_k).
\end{align*}
We will further discuss the justification for only letting the location parameter vary between samples in Section~\ref{sec:coarsening}. 

We assign multivariate Gaussian and inverse Wishart priors for the means and covariances of the kernels:
\[\bm\xi_{0,k}\sim N(\mathbf{b_0}, \mathbf{B_0}) \]
\[\bm\Sigma_k\sim \mathcal{W}^{-1}(m, \bm\Lambda)\]
and follow \cite{Liseo2013} and \cite{Parisi2018} in the assignment of a prior for the skewness parameter:
\[p(\bm\delta_k, \bm\Sigma_k)=p(\bm\delta_k\mid \bm\Sigma_k)p(\bm\Sigma_k) \]
\[p(\bm\delta_k\mid \bm\Sigma_k) = \left(\frac{\pi^{\frac p2}}{\Gamma(\frac p2 + 1)}  \sqrt{|\bm\Omega_k|}\right)^{-1}, \]
where $\bm\delta$ is a transformed version of $\bm\alpha$. The priors for $\bm\delta$ and $\bm\Sigma$ induce priors on an alternative equivalent parametrization in terms of $\bm\psi$ and $\bm G$, which are derived from $\bm\delta$ and $\bm\Sigma$ by  multiplying the following Jacobian term, defined separately for each cluster $k\in\mK$:
\[|\mathcal J_k[(\bm\xi_{0,k},\bm\Sigma_k,\bm\delta_k)\to(\bm\xi_{0,k},\bm G_k,\bm\psi_k)]|=\prod_{j=1}^p (\bm G_k(j,j)+\bm\psi_k(j)^2)^{-\frac12}.\]
In practice, we will infer on $\bm\psi_k$ and $\bm G_k$ and then transform them back to the original parameters. (See Supplementary Materials~\ref{sec:msn} for further details on the alternative parameterizations of the multivariate SN distribution and its computational advantages.)

\begin{figure}[!h]
\includegraphics[width=0.99\textwidth]{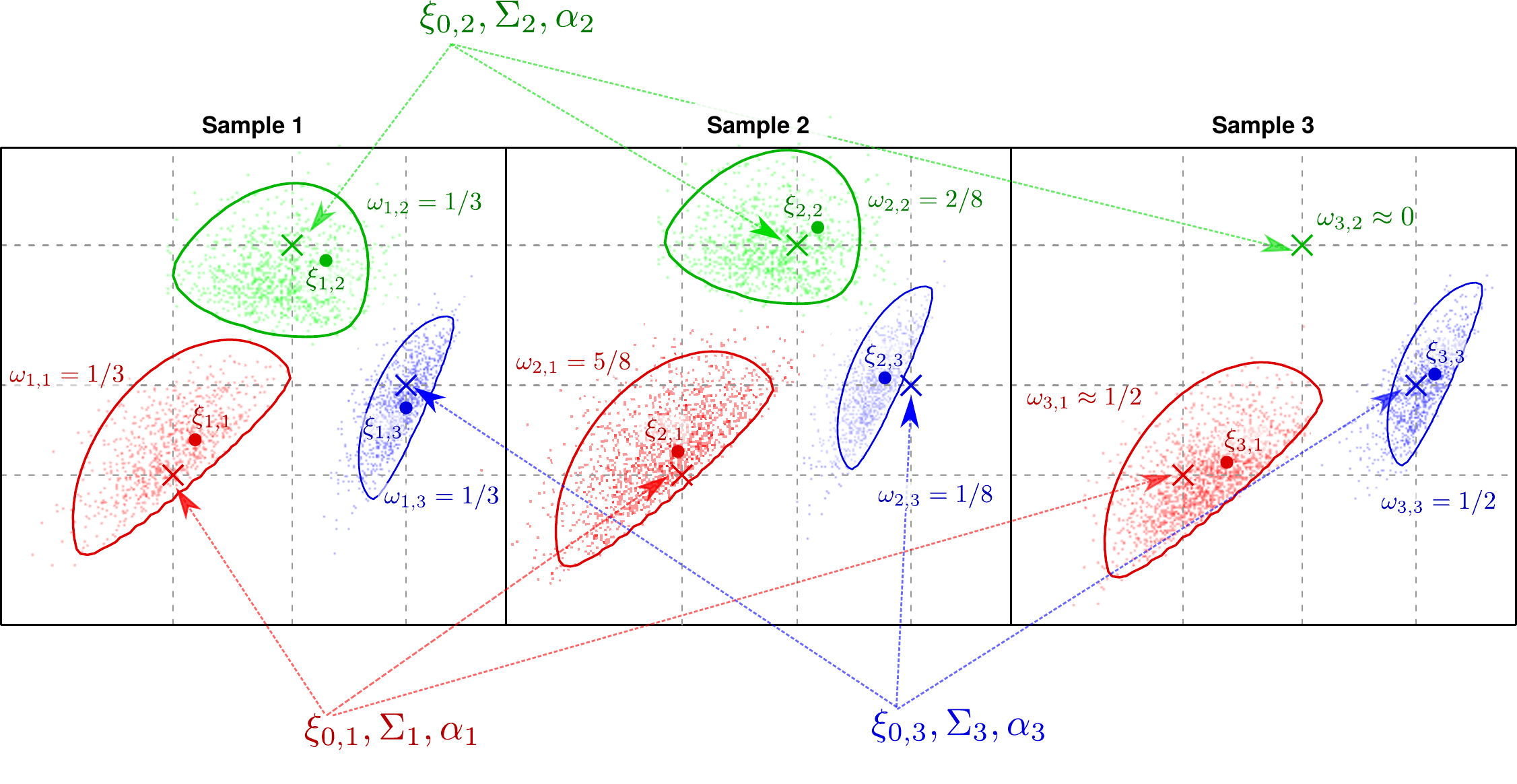}
\caption{An illustration of 3-sample data set generated from our model. The ``centroid'' cluster parameters $\bm\xi_{0,k}$, $\bm\Sigma_k$ and $\bm\alpha_k$ are all shared across the samples. Each sample has its own set of weights $\bm\omega_{j,k}$ and, e.g., cluster 2 in sample 3 is empty with $\omega_{3,2}\approx0$ so that $T_{i,3}\ne 2$ for all $i\in\{1,\hdots,n_3\}$. The location parameters of the first cluster $\bm\xi_{1,1}, \bm\xi_{2,1}$ and $\bm\xi_{3,1}$ are spread around $\bm\xi_{0,1}$. Similarly, $\bm\xi_{1,2}, \bm\xi_{2,2},\bm\xi_{3,2}$ are spread around $\bm\xi_{0,2}$ and $\bm\xi_{1,3},\bm\xi_{2,3}$ and $\bm\xi_{3,3}$ are spread around $\bm\xi_{0,3}$.\label{fig:illustrate_model}}
\end{figure}

We further assign an inverse-Wishart prior to the covariance of the normal distribution of the cluster means around the grand mean:
\[\bm E_k\sim \mathcal{W}^{-1}(\nu_0,\bm E_0).\]
This completes the specification of our hierarchical model. Figure~\ref{fig:MPSK} provides a graphical model representation of the full hierarchical model while Figure~\ref{fig:illustrate_model} illustrates the structure of the data generated from this model. 

\subsection{Model robustification through coarsening}
\label{sec:coarsening}
We use the ``coarsening'' strategy introduced in \cite{miller2018} to make inference robust to model misspecifications. 
This is particularly relevant in the context of FCM data - where clusters with parametric distributions usually do not perfectly fit the actual shape of cell subsets and the massive sample sizes of FCM data makes the resulting inference particularly sensitive to such model misspecifications.

Generally, for a model family indexed by some parameter $\theta$, one defines an ``idealized distribution'' as a member in the model family that we use to represent the data generative mechanism. (Here it is the full hierarchical model presented above.) 
The coarsening approach assumes the existence of unobserved ``ideal data'', say $Y_1,\hdots,Y_n$ from this distribution. The observed data, say $y_1,\hdots,y_n$, are assumed to be drawn from a true distribution which is in a $R$-neighborhood (under some discrepancy measure $d$) of the idealized distribution. When the observed and ideal distributions are the same, Bayesian inference is conducted on the standard posterior distribution, $p(\theta\mid Y_1=y_1,\hdots,Y_n=y_n)$. When they differ, Bayesian inference is performed on the ``coarsened posterior'': $p(\theta\mid d(\{Y_1,\hdots,Y_n\},\{y_1,\hdots,y_n\})<R)$. That is, the ``posterior'' is computed conditional on the event that the empirical distributions of the observed and ideal data are within a $R$-ball defined by a discrepancy measure $d$. $R$ can be taken as a random variable, and assigned a prior. When $R\sim\exp(\gamma)$ and $d$ is the Kullback-Leibler divergence (or $d_n$ a consistent estimator of it), the coarsened posterior is approximated by:
\[ \pi(\theta\mid d_n(\{Y_1,\hdots,Y_n\},\{y_1,\hdots,y_n\})<R) \approxprop \pi(\theta) \prod_{i=1}^n p_\theta (y_i)^{\zeta} \]
where $\approxprop$ means ``approximately proportional to'', $\zeta=\frac{\gamma}{\gamma+n}$, 
and $\pi(\theta)$ is the prior on $\theta$. 
Given some $\zeta\in[0,1]$, $\prod_{i=1}^n p_\theta (y_i)^{\zeta}$ is referred to as the ``power likelihood'' and $\pi(\theta)\prod_{i=1}^n p_\theta (y_i)^{\zeta}$ is referred to as the ``power posterior''. This form is easily implemented within the context of Gibbs sampling for mixture models, and we do so for our model. In our context, we consider the entire hierarchical DPM model as the idealized distribution. 
The types of deviations that the coarsening procedure tolerates depends on the discrepancy measure. The Kullback-Leibler divergence is well suited for FCM data as it is sensitive to changes in the shape of a distribution (as opposed to, say, moving probability mass to outlying regions). 
It allows the hierarchical kernel in our model to focus on the location alone, which simplifies the model and the corresponding sampling algorithm.

\subsection{Cross-sample calibration through posterior prediction}
\label{sec:calib}
To calibrate multiple samples, we aim to shift each observation in a cluster in each sample by the estimated difference between the grand ``centroid'' mean and the sample-specific mean of that cluster. That is, when $T_{i,j} = k$ we compute a corrected value for each observation by adjusting for the shift in the mean:
\begin{align*} 
\tilde{\bm y}_{i,j}&=\bm y_{i,j}-((\bm\xi_{j,k}+\bm\omega_k\bm\delta_k\sqrt{\nicefrac{2}{\pi}})-(\bm\xi_{0,k}+\bm\omega_k\bm\delta_k\sqrt{\nicefrac{2}{\pi}}))=\bm y_{i,j}-(\bm\xi_{j,k}-\bm\xi_{0,k}) 
\end{align*}
where $\tilde{\bm y}_{i,j}$ is the shifted observation corresponding to the original observation $\bm y_{i,j}$. In the above, $\bm\omega$ and $\bm\delta$ are alternative parameterizations to $\bm\Sigma$ and $\bm\alpha$. (See Supplementary Materials~\ref{sec:msn} for further details on the multivariate SN distribution.) To incorporate the posterior uncertainty in the cluster assignment, we follow the technique suggested in \cite{soriano2018arxiv} for calibration by integrating out the cluster assignment variables $T_{i,j}s$:
\begin{align*}
\E[\tilde{\bm y}_{i,j}\mid \{\bm y_{i,j}\}]&=\bm y_{i,j}-\E[\bm\xi_{j,k}-\bm\xi_{0,k}\mid  \{\bm y_{i,j}\}]\approx \bm y_{i,j}-\frac1N\sum_{n=1}^N (\bm \xi_{j,T_{i,j}^{(n)}}-\bm \xi_{0,T_{i,j}^{(n)}}).
\end{align*}
A desirable byproduct of this integrating-out strategy is that through it we also bypass potential label switching issue, which will be discussed further in the next subsection.

\subsection{Posterior computation by hybrid Gibbs-PMC sampling\label{sec:implementation}}

The multimodality of the SN likelihood poses difficulties to EM estimation in frequentist settings and to Gibbs samplers in Bayesian settings. For example, \cite{sylvia2010} offered a conjugate structure and a Gibbs sampler to perform Bayesian estimation of the parameters of the SN distribution. \cite{Liseo2013} demonstrated how this approach may fail when multimodality arises in the likelihood. 
In addition, our experimentation of the conjugate prior approach also suggest that huge amounts of data are required to overwhelm the prior and allow correct inference for highly skewed kernels. Furthermore, the conjugate prior structure entails the elicitation of a joint prior for the location parameter $\bm\xi$ and the skewness parameter $\bm\psi$, which in the current context of application to FCM data will make tasks such as cross-sample calibration difficult. For these reasons, it is important to seek a reasonable non-conjugate prior on the SN kernel parameters along with an efficient computational strategy in the context of FCM data.

\cite{Liseo2013} suggested utilizing the Population Monte Carlo (PMC) approach to tackle the problem of multimodality while allowing prior modeling on the location parameter independently from those of the other SN parameters---that is, $\bm\xi \indep (\bm\Sigma,\bm\delta)$ {\em a priori}. 
This strategy particularly suits our need as it allows us to select flexible and intuitive priors that offer a simple hierarchical structure for the location parameter as in our model. Because the resulting priors are not conjugate, we cannot use a vanilla Gibbs sampler. (In contrast, \cite{hejblum2017}, \cite{dundar2014}, and \cite{soriano2019ba} all constrained their models so that the prior structure will be conjugate and allow blocked Gibbs sampling.) We thus construct a hybrid ``Gibbs-PMC'' sampler that utilizes PMC moves for the SN parameters while using Gibbs moves on the rest of the model.

{\em A ``Gibbs-PMC' hybrid Sampler.}
Our sampler uses PMC moves for the SN parameters $\xi_{j,k}$, $G_k$, $\psi_k$, $\xi_{0,k}$, $E_k$ and $z_{i,j}$, and then given summary statistics (mean) of all particles for these parameters; it uses Gibbs moves for the other parameters parameters $\pi_{j,k}$ and $T$, and Metropolis-Hastings moves for just the DPM concentration parameter $a_\eta$. The full sampling algorithm is as follows.  
\begin{itemize}
\item Step 0: Initialize a population of $M$ particles $\xi_{j,k}^{1:M}$, $G_k^{1:M}$, $\psi_k^{1:M}$, $\xi_{0,k}^{1:M}$, $E_k^{1:M}$ and $z_{i,j}^{1:M}$.
\item Step $t>0$:
\begin{itemize}
\item Let $n_k$ be the number of observations in cluster $k$ and $n_{j,k}$ the number of observations for sample $j$ in cluster $k$.
\item Update the Dirichlet pseudo-count parameters $a_\eta$ using a Metropolis-Hastings step with the proposal: \[a^*_\eta\mid a_\eta\sim\mathrm{Gamma}(a_\eta^2 \cdot a_0, a_\eta \cdot a_0)\] where $a_0$ is calibrated during the burn-in iterations.
\item Sample from the full conditional of the mixture weights.
\item For $k$ in $1,2,\hdots,K$:
\begin{itemize}
\item Sample $M$ particles $z_1^{1:M},\hdots,z_{n_k}^{1:M}$ from the proposal $q^{(m)}_z$ as the full conditional distribution of $z_{i,j}$ (i.e. $z_{i,j}^{(m)}$ depends via $q^{(m)}_z$ on $\psi_k^{(m)}$, $G_k^{(m)}$, $\xi_{j,k}^{(m)}$ and $\xi_{0,k}^{(m)}$ for $m=1,\hdots,M$).
\item In a random order, perform the next 5 updating steps:
\begin{enumerate}
\item Sample $M$ particles $\xi_{j,k}^{1:M}$ from the proposal $q^{(m)}_{\xi_{j,k}}$ as the full conditional distribution of $\xi_{j,k}$.
\item Sample $M$ particles $G_k^{1:M}$ from the proposal $q^{(m)}_{G_k}$ as the inverse-Wishart part of the full conditional distribution of $G_k$.
\item Sample $M$ particles $\psi_{j,k}^{1:M}$ from the proposal $q^{(m)}_{\psi_k}$ as the $p$-dimensional multivariate normal part of the full conditional distribution of $\psi_{j,k}$.
\item Sample $M$ particles $\xi_{0,k}^{1:M}$ from the proposal $q^{(m)}_{\xi_{0,k}}$ as the full conditional distribution of $\xi_{0,k}$.
\item Sample $M$ particles $E_k^{1:M}$ from the proposal $q^{(m)}_{E_k}$ as the full conditional distribution of $E_k$.
\\

\textit{(For every $k$, if $n_k=0$ sample particles from priors.)}
\end{enumerate}
\end{itemize}
\item Compute the ratios \[\varrho^{(m)} \h\propto\h \frac{\pi\left(\xi_{j,k}^{(m)}, G_k^{(m)}, \psi_k^{(m)}, \xi_{0,k}^{(m)}, E_k^{(m)}, \{z_{i,j}^{(m)}\}\mid \{y_{i,j}\}\right)}{q^{(m)}\left(\xi_{j,k}^{(m)}, G_k^{(m)}, \psi_k^{(m)}, \xi_{0,k}^{(m)}, E_k^{(m)}, \{z_{i,j}^{(m)}\}\right)}\]
where $q^{(m)}$ is the joint proposal for each particle.
\item Scale the $\{\varrho^{(m)}\}$ to sum to 1.
\item Resample $\left\{\xi_{j,k}^{(m)}, G_k^{(m)}, \psi_k^{(m)}, \xi_{0,k}^{(m)}, E_k^{(m)} \right\}_{m=1,\hdots,M}$ according to the weights $\{\varrho^{(m)}\}$.
\item Compute (mean over $M$ index) $\overline{\xi_{j,k}^{1:M}}$, $\overline{G_k^{1:M}}$, $\overline{\psi_k^{1:M}}$, $\overline{\xi_{0,k}^{1:M}}$, $\overline{E_k^{1:M}}$ and $\overline{z_{i,j}^{1:M}}$. 
\item Sample from the full conditional distribution of $T_{i,j}$, based on the values $\overline{\xi_{j,k}^{1:M}}$, $\overline{G_k^{1:M}}$, $\overline{\psi_k^{1:M}}$ and $\{\pi_{j,k}\}$
\item (For each $t$) store: $\overline{\xi_{j,k}^{1:M}}$, $\overline{G_k^{1:M}}$, $\overline{\psi_k^{1:M}}$, $\overline{\xi_{0,k}^{1:M}}$, $\overline{E_k^{1:M}}$ and $\overline{z_{i,j}^{1:M}}$, $\{T_{i,j}\}$, $a$ and $\{\pi_{j,k}\}$.
\item Merge clusters for which the Kullback-Leibler divergence is smaller than a preset threshold.
\end{itemize}
\end{itemize}
The full conditional and proposal distributions used in the sampler are described in Supplementary Materials~\ref{sec:conditional}.

In implementing the sampler, we utilize the finite-dimensional symmetric Dirichlet distribution approximation \citep{Ishwaran2001} to the Dirichlet process mixture. With this approximation, the $J$ infinite sequences of mixture weights $\{\omega_{j,k}\}_{k\in\mK}$ are replaced for each $j=1,\hdots,J$ by:
\[\{\omega_{j,k}\}_{k\in\mK}\sim \mathrm{Dirichlet}(\eta/K, \hdots, \eta/K)\]
where we need to choose the maximal number of clusters $K$. 
A known issue of this approximation for DPM when applied to large data sets is that the posterior will concentrates on a diverging number of clusters and $K$ is often the \textit{de facto} number of estimated clusters. In our approach, however, this issue is addressed with the coarsening strategy---we can set $K$ to be very large while the estimated number of clusters will still be much smaller than $K$ even for large and noisy data sets such as those from FCM.  

A common problem in inference algorithms for mixture models that utilize a cluster assignment variable is known as label switching. Since the values the cluster assignment variable assigns (the labels) to the different clusters are exchangeable, the prior and posterior distributions for the parameters of the mixture components are symmetric with respect to permutations of the labels. This problem does not arise in our framework when calibration is the sole purpose as our calibration strategy integrates out the different labeling scenarios. For cell classification, on the other hand, label switching needs to be addressed and many possible strategies are available. Our software implementation handles this issue \textit{post hoc} using the method of \cite{cron2011}. A coherent classification is maintained by choosing a reference classification, which we take from the last iteration of the MCMC chain. Then, for each saved classification a cost matrix is computed (based only on the values of the cluster assignment variable) and minimized by selecting a permutation on its columns using the Hungarian algorithm of \cite{munkres1957}. The resulting minimizing permutation is then applied to the cluster labels. 

\section{Simulation study\label{sec:nmrcl}}
\begin{figure}[ht]
\centering
\includegraphics[width=0.7\textwidth]{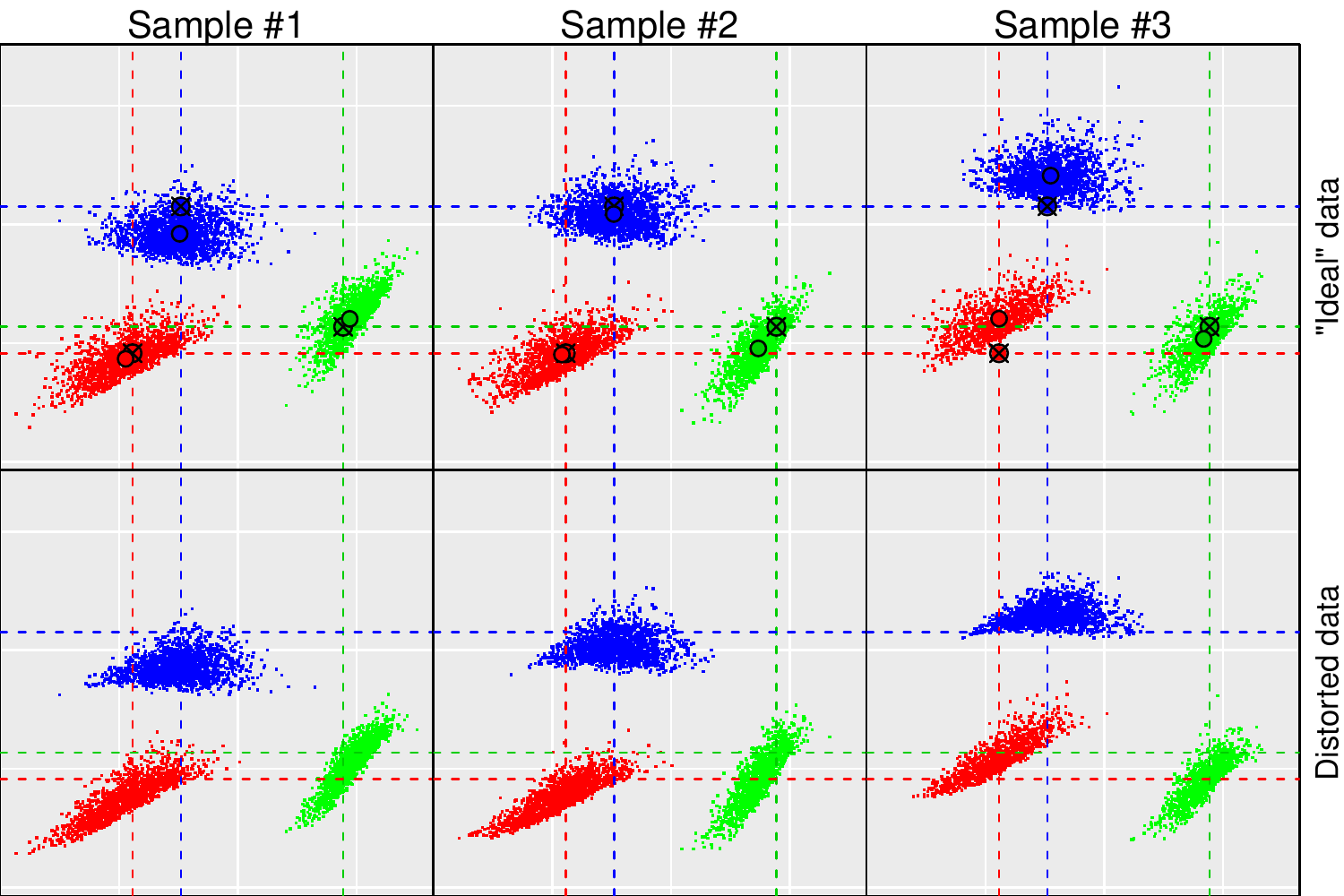}
\caption{Simulated data. Top row: an ``ideal'' data set in 2 dimensions with 3 clusters in 3 samples. Bottom row: a distorted version of the ``ideal'' data.
The colors correspond to true cluster labels. The circles designate true sample-specific cluster means. The crossed circles designate true grand means across the samples.\label{fig:sim_data}}
\end{figure}

We demonstrate our method using two-dimensional simulated data for ease of visualization. To examine the robustness of our method to model misspecification in kernel shapes and the effects of coarsening, we apply our method to two different data sets. The first is simulated under a finite mixture model with 3 bivariate skew-normal components, with the same hierarchical structure for the kernels as in our model. That represents a case our model is correctly specified even without coarsening. The second data set is generated by ``distorting'' the first, ``ideal'', data set to induce model misspecification. In the distorted samples each cluster is narrowed asymmetrically (see supplement for source code for generating these data). 
Figure~\ref{fig:sim_data} presents both the ``ideal'' data (top row) and the ``distorted'' data (bottom row). 

We apply COMIX with a maximal number of 150 clusters to each data set. In order to examine the effects of coarsening, we show model estimates and calibration results with coarsening ($\zeta=0.2$) and without it ($\zeta=1$). We have found in many numerical studies that $\zeta=0.2$ provides robust results for classification and calibration in many contexts where the observed data deviate from the theoretical skew kernels. As such we adopt 0.2 as a default value in our software. In applications we recommend the user to still carry out a context-specific sensitivity analysis for $\zeta$ to ensure that an appropriate value is selected. We provide such an example in our later case studies.  

\begin{figure}[!h]
\centering
\includegraphics[width=0.7\textwidth]{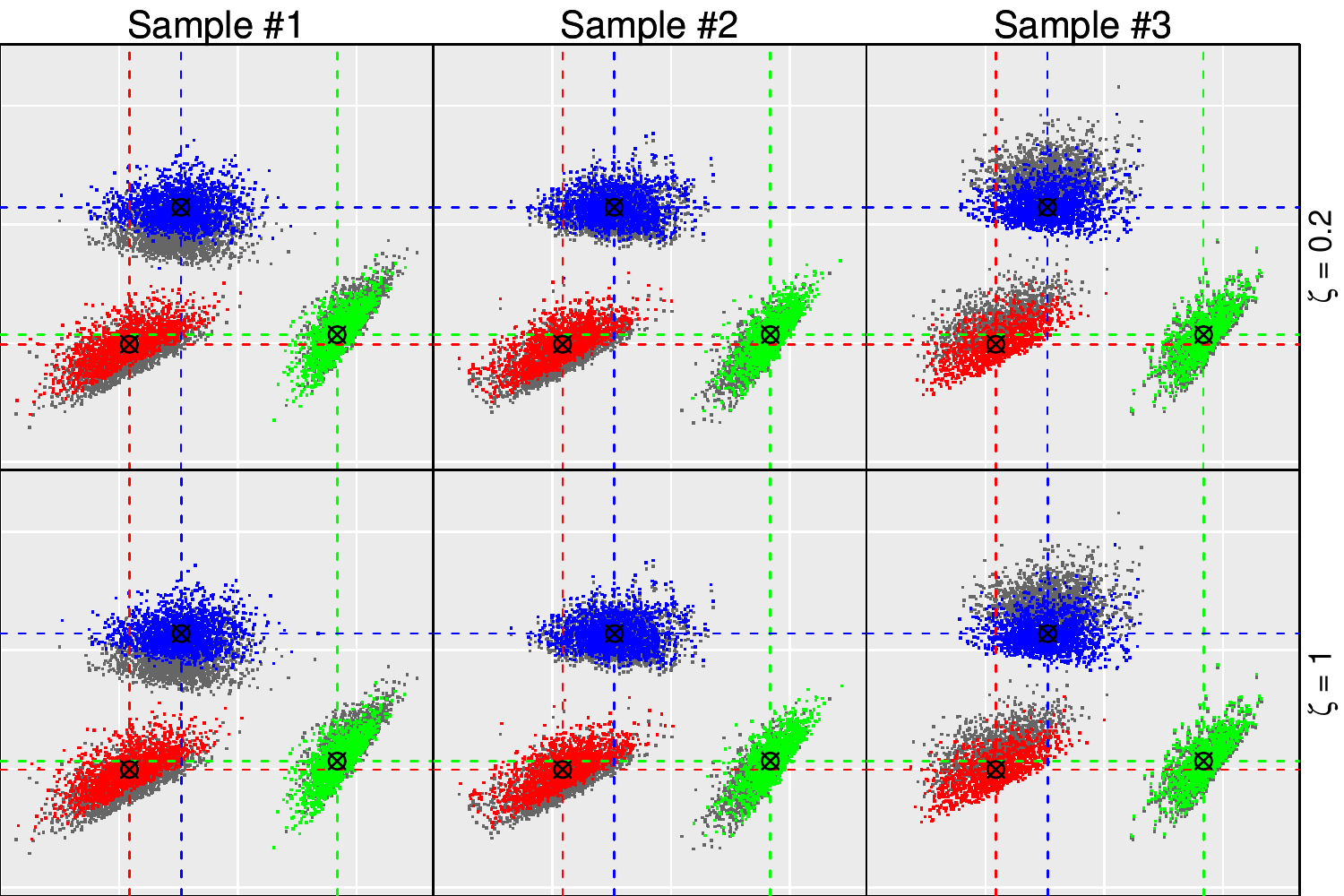}
\caption{Calibrated samples for the ``ideal'' data with coarsening (top row, $\zeta=0.2$) and without (bottom row, $\zeta=1$).  Dark grey: the observations in the ``ideal'' data prior to calibration. The colors indicate estimated cluster label assignment. The crossed circles indicate estimated grand cluster means.\label{fig:2d_calib_base}}
\end{figure}

\begin{figure}[!h]
\centering
\includegraphics[width=0.7\textwidth]{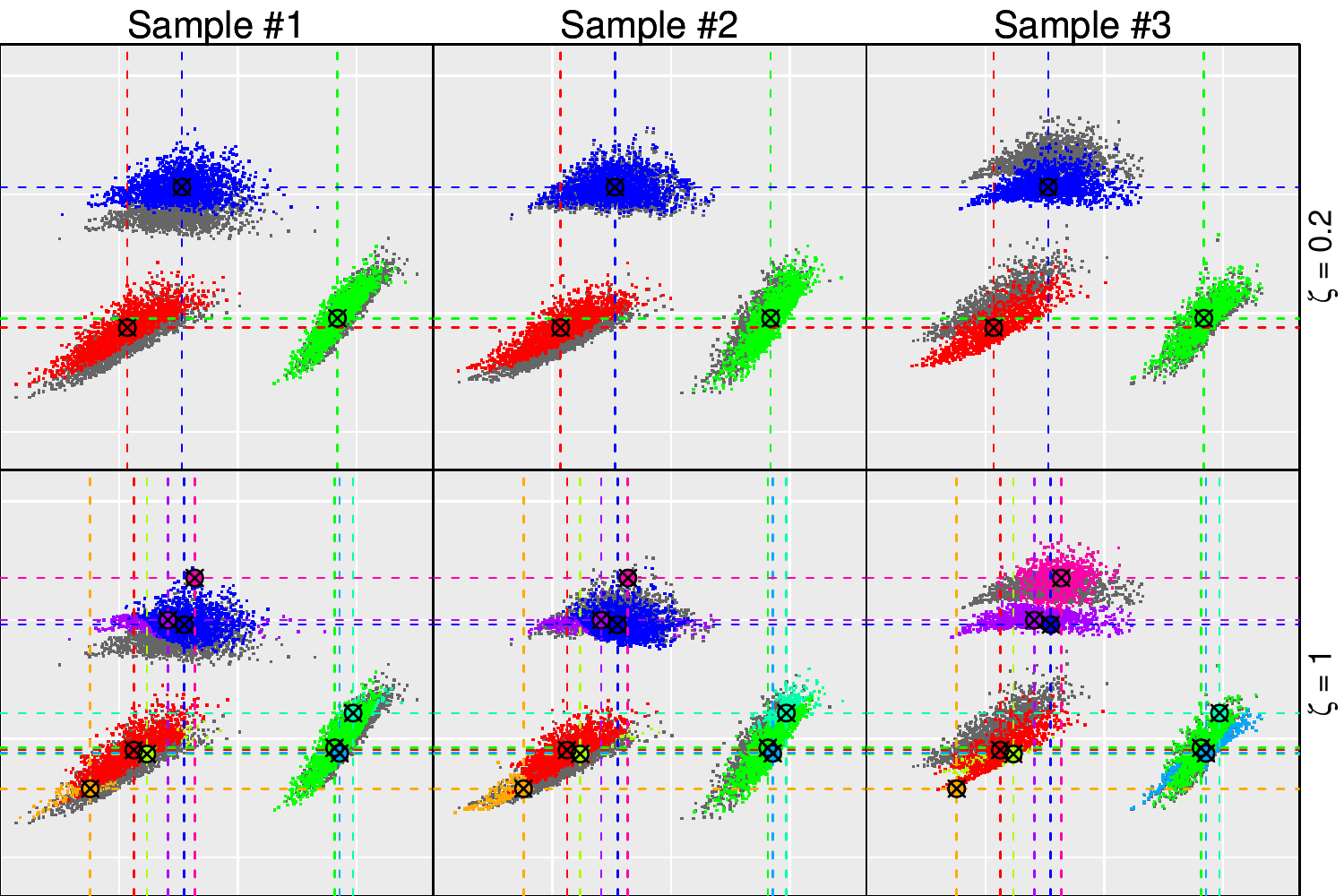}
\caption{Calibrated samples for the distorted data with coarsening (top row, $\zeta=0.2$) and without (bottom row, $\zeta=1$). Dark grey: the observations in the ``distorted'' data prior to calibration. The colors indicate estimated cluster label assignment. The crossed circles indicate estimated grand cluster means.\label{fig:2d_calib}}
\end{figure}

The classification and calibration results for the data drawn from the ``ideal'' kernels are shown in Figure~\ref{fig:2d_calib_base} and the results for the distorted data are shown in Figure~\ref{fig:2d_calib}. In Figure~\ref{fig:2d_calib_base}, where the clusters are drawn from the same distribution as in our generative model, each true cluster is identified as such both with and without coarsening. However, in Figure~\ref{fig:2d_calib}, where the distorted data are not exactly from SN kernels as our generative model prescribes, the true clusters are broken into several clusters without coarsening ($\zeta=1$). Coarsening helps identify and align the distorted clusters correctly even though they are misspeciefied by our generative model.

\vfill

\section{Case study: the EQAPOL data}
\label{sec:casestudies}
\subsection{A 6-dimensional data set}
\label{sec:flow}
We apply our method to FCM samples processed in 10 different laboratories participating in the External Quality Assurance Program Oversight Laboratory (EQAPOL) program. In this study, peripheral blood samples from the \emph{same} subject was sent to different laboratories together with lyophilized antibody panels for processing. Because the samples are from the same individual and subjected to the same sample preparation protocol, the analytical results should be almost identical in principle. We demonstrate our calibration method to the 10 samples with six features (FSC, SSC, CD3, CD4, CD8, multiplexed IFN-$\gamma$ + IL-2).  

\begin{figure}[h!]
\centering
\includegraphics[width=1\textwidth]{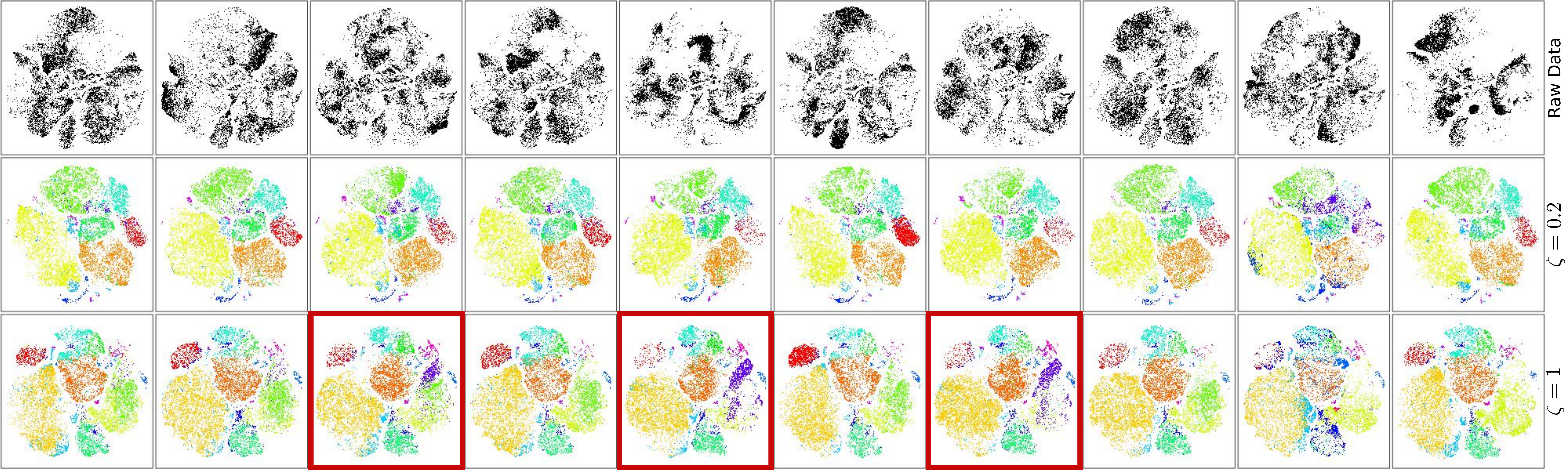}
\caption{t-SNE plots for the 6-dimensional data. Top row: raw data. Middle row: calibrated data for four samples with $\zeta=0.2$. Bottom row: calibrated data for four samples with $\zeta=1$.  Color coding corresponds to estimated cluster assignment variable. The three labs highlighted with red boundary demonstrate where the classification and calibration for $\zeta=1$ is inferior to that of $\zeta=0.2$.\label{fig:tsne_flow}}
\end{figure}

Figure \ref{fig:tsne_flow} shows the t-SNE plots \citep{Maaten2008} for the raw and calibrated data for the 10 labs. As in the previous section, we show the results for $\zeta=0.2$ and $\zeta=1$. 

\begin{figure}[h!]
\centering
\includegraphics[width=1\textwidth]{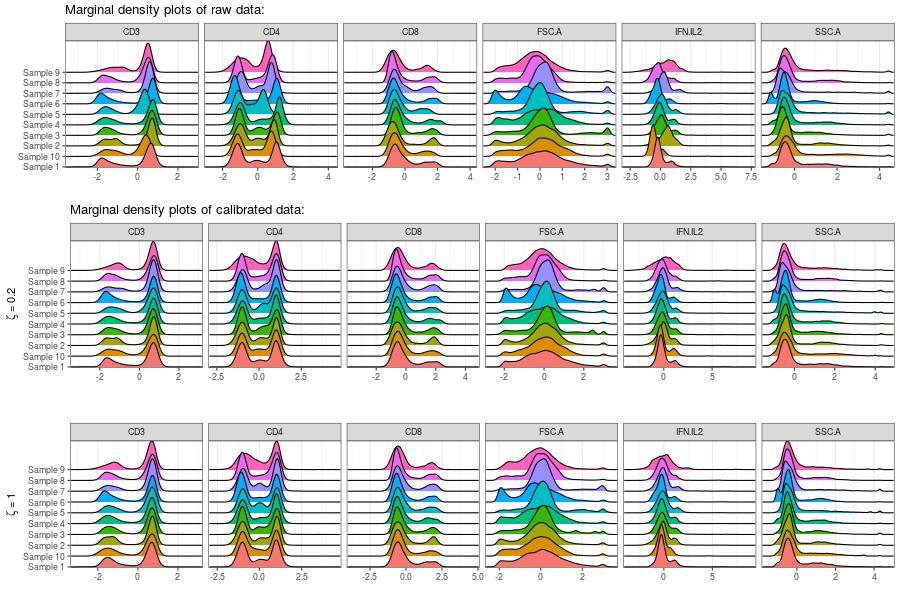}
\caption{Marginal density plots for all markers and samples in the 6-dimensional FCM data set.\label{fig:density_flow}}
\end{figure}

The aligned data are much more similar across samples, and both large and small clusters are properly aligned. Specifically, when $\zeta=0.2$, in all samples the number of estimated clusters is 11, in accordance with the underlying assumption for the batch control study that all samples are the same and thus have the same number of clusters. When $\zeta=1$ the number of estimated clusters (that is, the number of unique values in the estimated cluster assignment variable) in the different samples varies between 13 to 15. This suggests that coarsening here is indeed useful in addressing model misspecification in the context of FCM data. We carry out a sensitivity analysis on the value of $\zeta$ and provide the details in Supplementary Materials~\ref{sec:sensitivity}. Finally, Figure~\ref{fig:density_flow} presents the marginal densities of the six markers before and after calibration.

\subsection{A 19-dimensional data set}
\label{sec:mass}
We further apply our method to a publicly available higher dimensional data set collected using a mass cytometer. 
We pre-processed the data with an Arcsinh transform using per channel parameters in the binary FCS file. Following Figure 2B in \cite{Kleinsteuber2016} we remove from the samples a ``spike-in'' batch control based on expression levels of the CD45 barcodes. We then extract the 19 markers described in \cite{Kleinsteuber2016} and demonstrate the calibration and classification results for Patient \#1 batch control across the 3 experimental settings.

\begin{figure}[ht]
\centering
\includegraphics[width=0.45\textwidth]{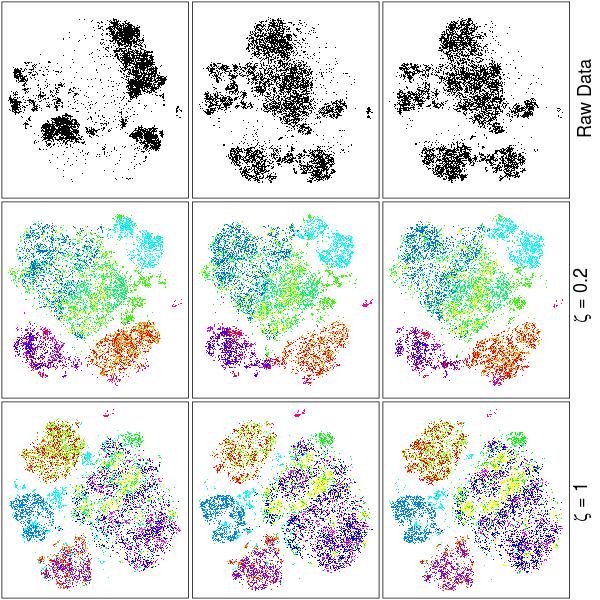}
\caption{t-SNE plots for the 19-dimensional data. Top row: raw data. Middle row: calibrated data for four samples with $\zeta=0.2$. Bottom row: calibrated data for four samples with $\zeta=1$.  Color coding corresponds to estimated cluster assignment variable.\label{fig:tsne_mass}}
\end{figure}

\begin{figure}[p]
\centering
\includegraphics[width=0.49\textwidth]{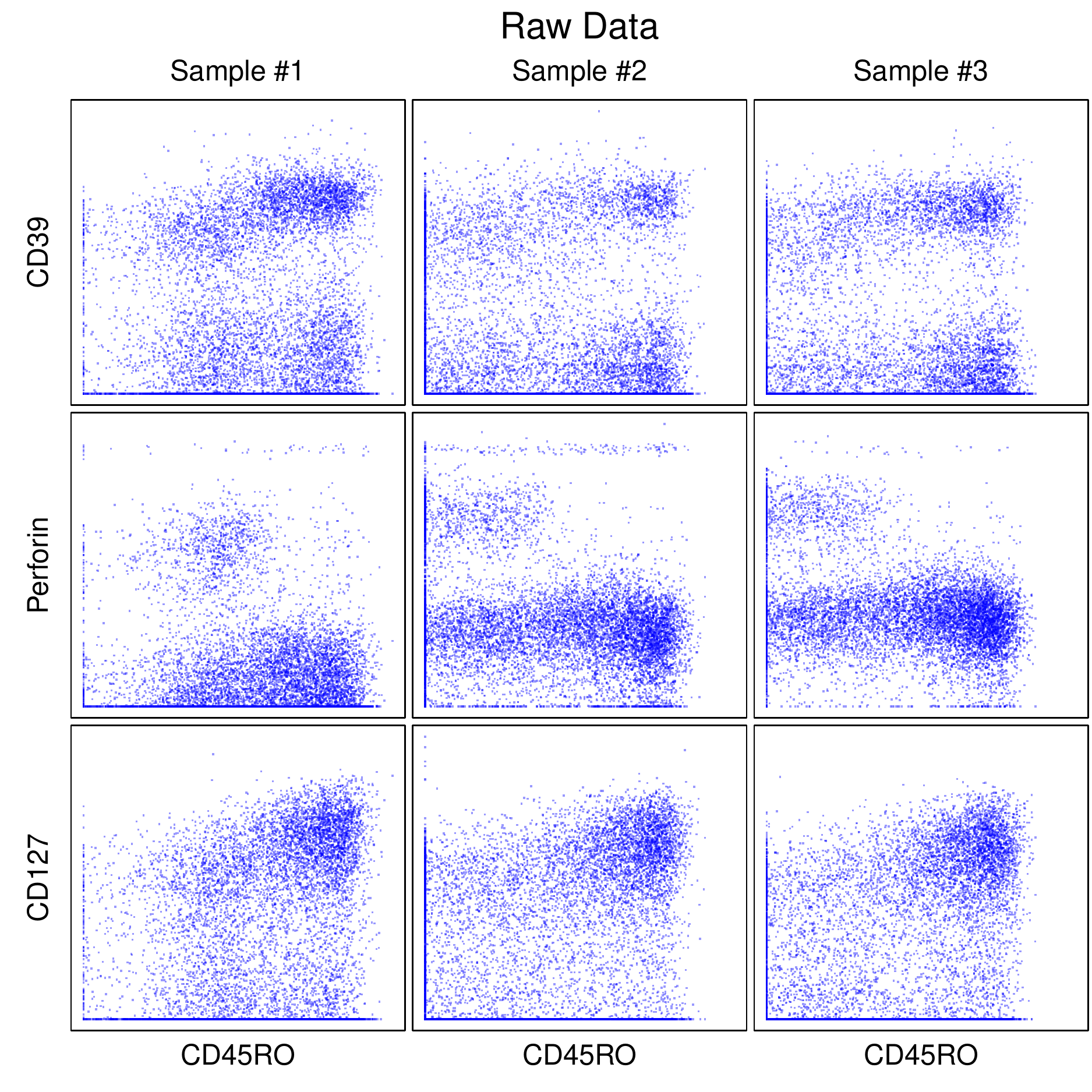}\\
\includegraphics[width=0.49\textwidth]{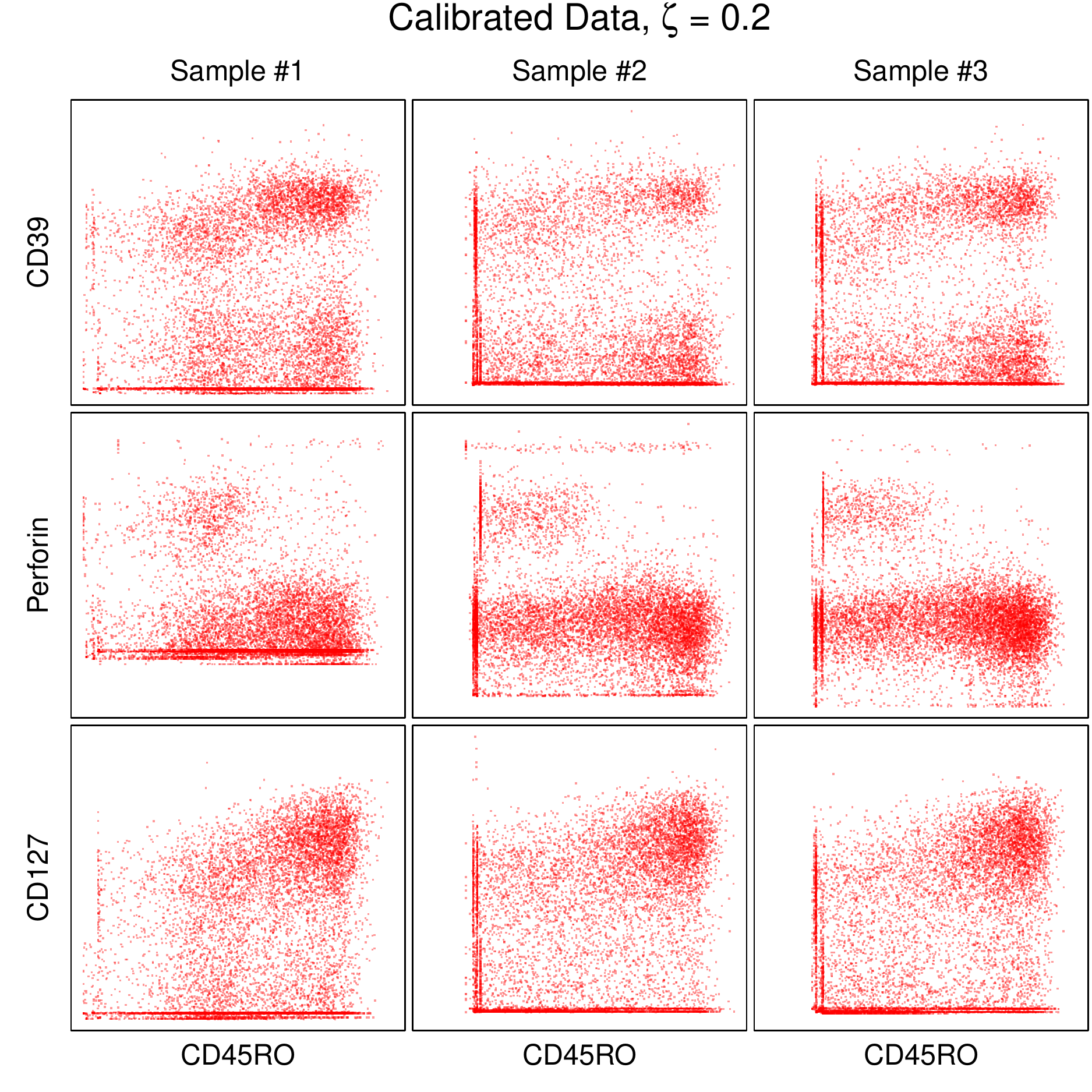}
\includegraphics[width=0.49\textwidth]{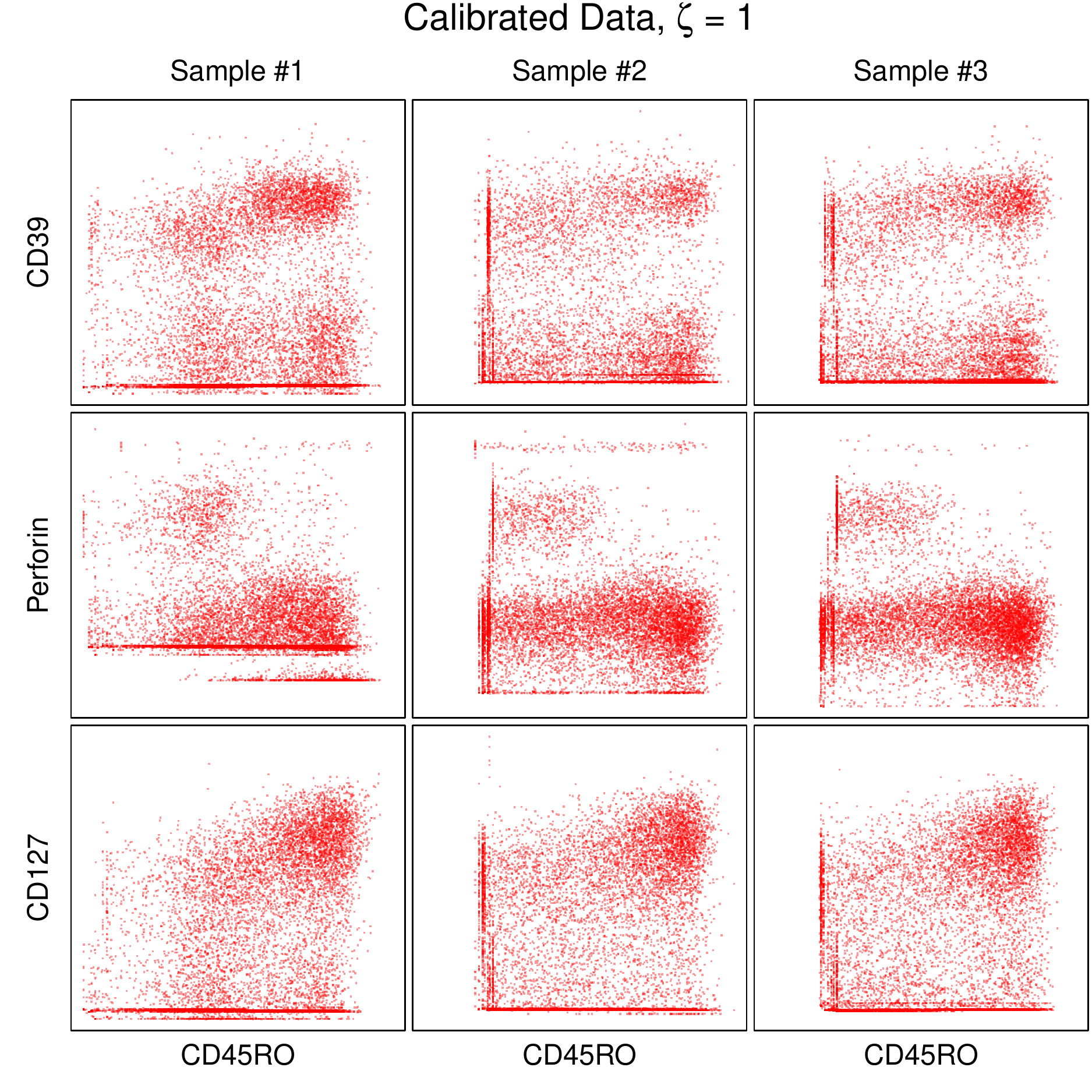}
\caption{Marginal scatter plots for some markers and all samples in high dimensional FCM data. Blue: raw data. Red: calibrated data for $\zeta=0.2$ and $\zeta=1$. \label{fig:margins_mass}}
\end{figure}

Figure \ref{fig:tsne_mass} shows t-SNE plots for the raw and calibrated data the of 3 samples for $\zeta=1$ and $\zeta=0.2$. 
Here too, the calibrated data is very similar across the samples, attaining our goal. In this case, we get that for $\zeta=1$ the number of estimated clusters is 12 across all samples, whereas for $\zeta=0.2$ it is 10. Figure~\ref{fig:margins_mass} shows some marginal scatter plots of the raw and calibrated data for the two settings. Although the results are largely similar, here too the smaller number of estimated clusters due to coarsening is beneficial, as the lowest cluster in the CD45RO-Perforin plot of the left sample for $\zeta=1$ appears to be artificial and uninformative.

The results for these two data sets show that COMIX works well for calibration of flow and mass cytometry data from 6 to 19 dimensions, which span the range of dimensions used in the vast majority of such experimental data sets. Calibration is broadly useful not just for multi-center studies, but also for studies across batches of data as cytometer performance characteristics, antibody lots, and sample preparation often vary over time, necessitating time-consuming and error-prone manual adjustment of gates across batches in manual analysis, and decreasing the robustness of automated methods that ignore the need for calibration. Hence, COMIX has broad application for all but the smallest flow and mass cytometry experimental data sets.

\section{Discussion}
We have presented a principled probabilistic approach for calibrating and classifying multi-sample FCM data. Our approach utilizes a flexible Bayesian nonparametric mixture model with multivariate SN kernels to incorporate the key features of FCM data are potentially high-dimensional and have massive sample sizes, and incorporate the ``coarsening'' strategy to make inference robust to model-misspecification. Moreover, we constructed a Gibbs-PMC hybrid sampler, which embeds PMC move for the SN parameters into a Gibbs sampler, thereby addressing the multimodality of the posterior on the kernel parameters.  

As described in our first case study, we applied the method to data from the EQAPOL program. This program is funded by National Institute of Allergy and Infectious Disease (NIAID) to improve the immune monitoring proficiency of national and international laboratories participating in NIAID-funded HIV clinical research. The current bottleneck for the EQAPOL flow cytometry proficiency testing program is the need for the central laboratory to identify deviations of the site gating strategy from a reference or consensus strategy for the purpose of feedback and remediation of site analysts. Currently, this is achieved by centralized manual re-analysis of the data sets from every participating site, an extremely labor-intensive and time-consuming activity. We are currently developing a semi-supervised processing pipeline that combines the calibration of data sets using COMIX with programmatic extraction of the gating strategy from FlowJo (the standard software used for manual analysis) workspace XML files to automatically report deviations of the gating strategy from the consensus or reference. We expect that this semi-supervised processing will eliminate the need for manual re-analysis but still provide the EQAPOL program with the information needed to provide site remediation effectively and allow program expansion to include more sites by removing the manual re-analysis bottleneck.

While our method is motivated by and developed for the purpose of analyzing multi-sample FCM data, the modeling and inference techniques used are generally applicable to other multi-sample data that can be effectively modeled by mixtures. In particular, the idea of adopting a flexible kernel, allowing hierarchical structure on the kernel, and incorporating coarsening to further robustify the method can all be readily applied to other types of data. 

\section*{Software}
Our R package {\tt COMIX} is available at \url{https://github.com/MaStatLab/COMIX}. Code for our numerical examples is available at \url{https://github.com/MaStatLab/COMIX_Numerical_Examples}.

\section*{Acknowledgment}
This research is supported by National Science Foundation Grant DMS-1749789. This work has also been supported in part through an EQAPOL collaboration with federal funds from the National Institute of Allergy and Infectious Diseases, National Institutes of Health, Contract Number HHSN272201700061C.

\bibliography{arxiv.bib}
\bibliographystyle{apalike}

\newpage
\beginsupplement
\setcounter{page}{1}
\section*{Supplementary Materials}

\section{Multivariate Skew Normals\label{sec:msn}}
\cite{azzalini1985} defined a class of distributions that generalize the normal distribution to non-symmetric extensions, and named this class as the skew normal (SN) distribution. \cite{azzalini1996} further extended this class to the multivariate case, and this formulation is the one we use in this work.

\subsection{First parametrization: $(\xi,\Sigma,\alpha)$}
A random vector $\bm X$ is said to have a $p$-dimensional standard SN distribution with correlation matrix $\bm\Omega_{p\times p}$ and skewness (or shape) parameter $\bm\alpha$ ($p$-dimensional column vector) when its density function is of the form 
\[f(\bm x;\bm0, \bm\Omega, \bm\alpha) = 2\varphi_p(\bm x;\bm0, \bm\Omega)\Phi_1(\bm\alpha^T \bm x)\]
where $\varphi_p$ is the density function of a $p$-dimensional multivariate normal with mean $\bm0$ and covariance $\bm\Omega$ and $\Phi_1$ a one dimensional standard normal CDF. To generalize, let $\bm\xi$ be a $p$-dimensional column vector and $\bm\omega=\mathrm{diag}(\Omega_{11}^{\nicefrac{1}{2}},\hdots,\Omega_{pp}^{\nicefrac{1}{2}})$ a diagonal matrix with the marginal scale parameters so that $\bm\Sigma=\bm\omega\bm\Omega\bm\omega$ represents the scale matrix. Then $\bm Y=\bm\xi+\bm\omega \bm X$ has a $p$-dimensional SN distribution, $SN_p(\bm\xi, \bm\Sigma, \bm\alpha)$, with density:
\[f(\bm y; \bm \xi, \bm \Sigma, \bm \alpha) = 2\varphi_p(\bm y; \bm \xi, \bm \Sigma)\cdot \Phi_1(\bm \alpha^T\bm\omega^{-1}(\bm y-\bm\xi))  \]

\subsection{Second parametrization: $(\xi,\Sigma,\delta)$ and a latent $Z$}
We follow the notation and results in \cite{Liseo2013} that introduce the useful latent structure of the SN distribution recognized earlier in \cite{azzalini1996}.
Given $\bm\Omega$ a correlation matrix (and the associated $\bm\omega$ and $\bm\Sigma$) and $\bm\alpha$ a skewness vector as before, let
$\bm\delta=(1-\bm\alpha^T\bm\Omega\bm\alpha)^{-\frac12}\bm\Omega\bm\alpha$ and define:

\[
\left(\begin{matrix}Z\\ \bm W\end{matrix}\right)\sim N_{p+1}\left[\left(\begin{matrix}0\\\bm{0}\end{matrix}\right), 
\left(\begin{matrix}1&\bm\delta^T\\\bm\delta&\bm\Omega\end{matrix}\right)\right]\]
and
\[\bm U = \begin{cases}\bm W&Z\geq0\\-\bm W&Z<0 \end{cases}\]
Then the random vector $\bm Y=\bm\omega \bm U+\bm\xi\sim SN_p(\bm\xi,\bm\Sigma, \bm\alpha)$. 
In addition, the joint density of $(\bm Y,Z)$ is given by
\[f_{p+1}(\bm y,z)=f_p(\bm y\mid z)f(z)=\varphi_p(\bm y; \bm \xi+\bm\omega\bm\delta|z|,\bm\omega(\bm\Omega-\bm\delta\bm\delta^T)\bm\omega)\cdot \varphi_1(z; 0,1)\]
In particular:
\[[\bm Y\mid Z=z] \sim \begin{cases}
N_p(\bm\xi+\bm\omega\bm\delta z, \bm\omega(\bm\Omega-\bm\delta\bm\delta^T)\bm\omega)&z\geq0\\
N_p(\bm\xi-\bm\omega\bm\delta z, \bm\omega(\bm\Omega-\bm\delta\bm\delta^T)\bm\omega)&z<0 \end{cases}\]
and 
\[\E(\bm Y)=\bm\xi+\bm\omega\bm\delta\sqrt\frac2\pi \]
\subsection{Third parametrization: $(\xi, G,\psi)$, $Z$ and augmented likelihood}
Let now $\bm\psi=\bm\omega\bm\delta$ and $\bm G=\bm\Sigma-\bm\psi\bm\psi^T=\bm\omega(\bm\Omega-\bm\delta\bm\delta^T)\bm\omega$. Then given observations $\bm y_1, \hdots ,\bm y_n$ with associated $z_1,\hdots,z_n$ the augmented likelihood is now:
\[\mathscr{L}((\bm\xi,\bm G,\bm\psi); \bm y_1,\hdots,\bm y_n,z_1,\hdots,z_n) = \prod_{i=1}^n \left\{\varphi_p(\bm y_i; \bm\xi+\bm\psi|z_i|, \bm G)\cdot \varphi_1(z_i; 0, 1)  \right\}  \]

In our estimation of the multivariate SN parameters, we follow the approach suggested \cite{Liseo2013} and use the parametrizations alternatively, multiplying by a Jacobian term where needed:
\[|\mathcal J[(\bm\xi,\bm\Sigma,\bm\delta)\to(\bm\xi,\bm G,\bm\psi)]|=\prod_{j=1}^p (\bm G(j,j)+\bm\psi(j)^2)^{-\frac12}\]

\section{Full Conditionals and MCMC Proposal\label{sec:conditional}}
The MCMC sampling is based on the following full conditional distributions, where $\zeta$ is the tuning parameter of the power likelihood.
\begin{enumerate}
\item Latent cluster assignments for $i=1,\hdots,n_j$ and $j=1,\hdots,J$:
\[\Pb(T_{i,j}=k\mid\hdots)\h\propto\h \pi_{j,k} \cdot SN(y_{i,j}; G_k, \xi_{j,k}, \psi_k)\]
\item Mixture weights:
\[[\pi_{j,1},\hdots,\pi_{j,K}\mid\hdots]\sim\operatorname{Dirichlet}(\zeta \cdot n_{j,1}+a/K,\hdots,\zeta \cdot n_{j,K}+a/K) \]

\item Latent random effects $z_{i,j}$:
\[f(z_{i,j}\mid T_{i,j}=k,\hdots)=\begin{cases}\varphi(z_{i,j}-m_{j,k}(i), v_k) & z_{i,j}\geq0\\ \varphi(z_{i,j}+m_{j,k}(i), v_k) & z_{i,j}> 0 \end{cases} \]
Where \[v_k=(1+\zeta\cdot \psi_k^TG_k^{-1}\psi_k)^{-1}\] and $m_{j,k}(i)$ is the $i$th element of the vector \[m_{j,k}=v_k(\zeta\cdot(y_{\cdot,j}^k-\mathbf{1}_n\otimes \xi_{j,k})^T G_k^{-1}\psi_k)\]
where $y_{\cdot,j}^k$ is the vector of all $n_j$ observations in sample $j$ and cluster $k$.

\item The scale parameter:
\[p(G_k\mid \hdots)\h\propto\h p(G_k)\cdot|\mathcal{J}_k|\cdot\mathcal{W}^{-1}(n_k+m, \Lambda_\star)\]
Where
\[\Lambda_\star = \Lambda + \zeta\cdot\sum_{i,j: T_{i,j}=k} (y_{i,j}-\psi_k|z_{i,j}|-\xi_{j,k})(y_{i,j}-\psi_k|z_{i,j}|-\xi_{j,k})^T  \]
(proposal: the inverse Wishart)

\item The skewness parameter:
\[p(\psi_k\mid \hdots)\h \propto \h p(\delta_k\mid\Sigma_k)|\mathcal{J}_k|\cdot\varphi_p\left(\psi_k; \frac{\sum_{i,j: T_{i,j}=k}|z_{i,j}|(y_{i,j}-\xi_{j,k})} {\sum_{i,j: T_{i,j}=k} z_{i,j}^2}, \frac{G_k}{\zeta\cdot\sum_{i,j: T_{i,j}=k} z_{i,j}^2}  \right)   \]
(proposal: the $p$-dimensional multivariate normal)

\item Group locations:
\begin{align*}
p(\xi_{j,k}\mid \hdots)\sim N_p\left(E^*(E_k^{-1}\psi_{0,k}+\zeta\cdot n_{j,k}G_k^{-1}(\overline{y}_{j,k}-\psi_k\overline{|z_{i.j}|}), E^*\right)
\end{align*}
Where $E^*= \left(E_k^{-1}+\zeta\cdot n_{j,k}G_k^{-1}\right)^{-1}$.
\item Grand locations:
\[\xi_{0,k}\mid \hdots \sim N_p(B^*(B_0^{-1}b_0+\zeta\cdot JE_k^{-1}\overline{\xi}_{j,k}), B^*)\]
Where 
\[B^* = \left(B_0^{-1}+\zeta\cdot JE_k^{-1}\right)^{-1}\]

\item Dispersion around grand location:
\[[E_k\mid \hdots]\sim\mathcal{W}^{-1}\left(\nu_0+J, (E_0^{-1}+\sum_j S_{j,k})^{-1}\right) \]
\end{enumerate}

\section{EQAPOL data: sensitivity analysis\label{sec:sensitivity}}
\begin{figure}[!h]
\centering
\includegraphics[width=0.8\textwidth]{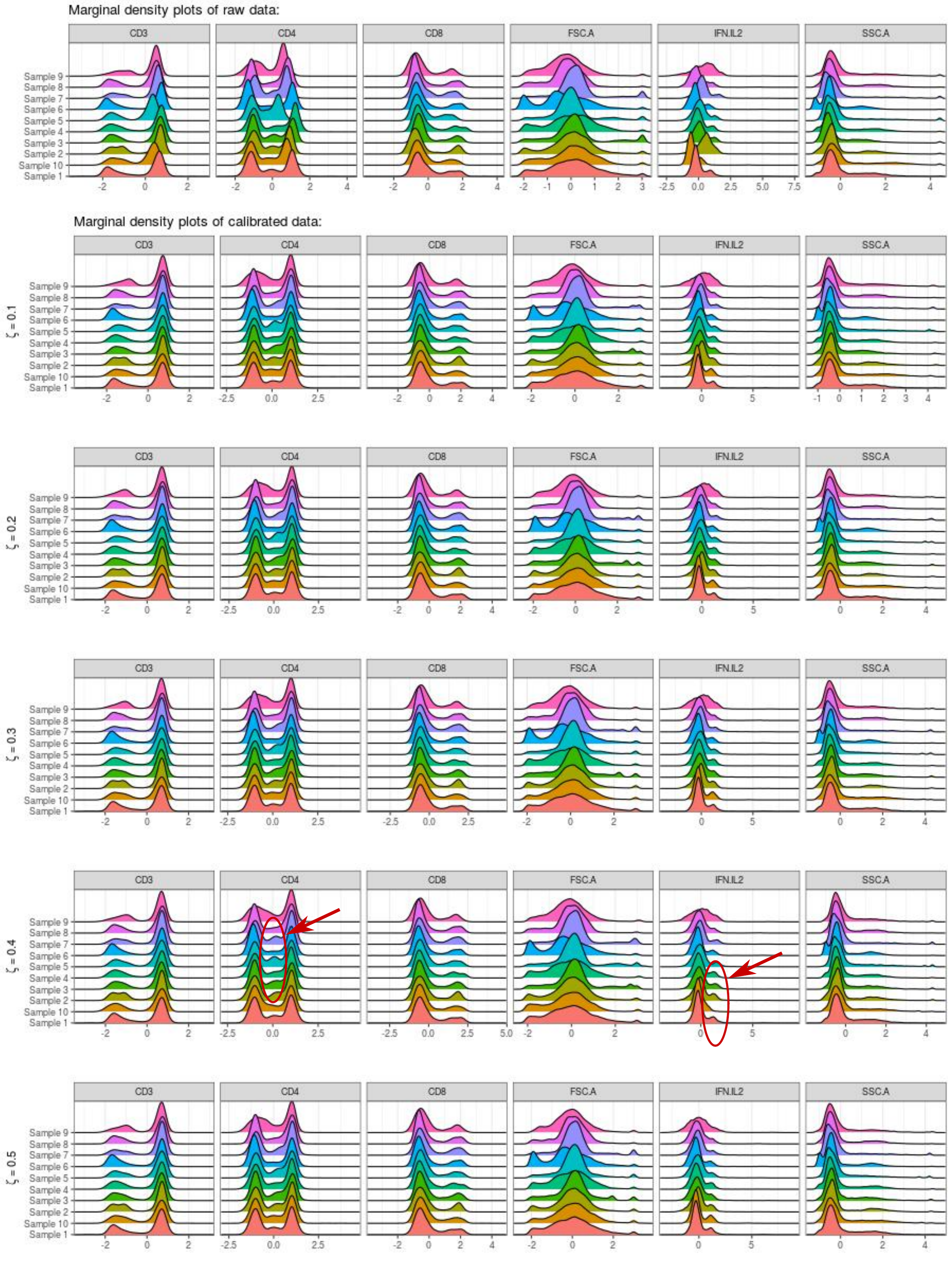}
\caption{Marginal density plots for all markers and samples in the 6-dimensional FCM data set. In red circles we highlight areas in some of the calibrated samples that are not as well aligned compared to the other calibrated samples.\label{fig:sensitivity1}}
\end{figure}

\begin{figure}[!h]
\centering
\includegraphics[width=0.8\textwidth]{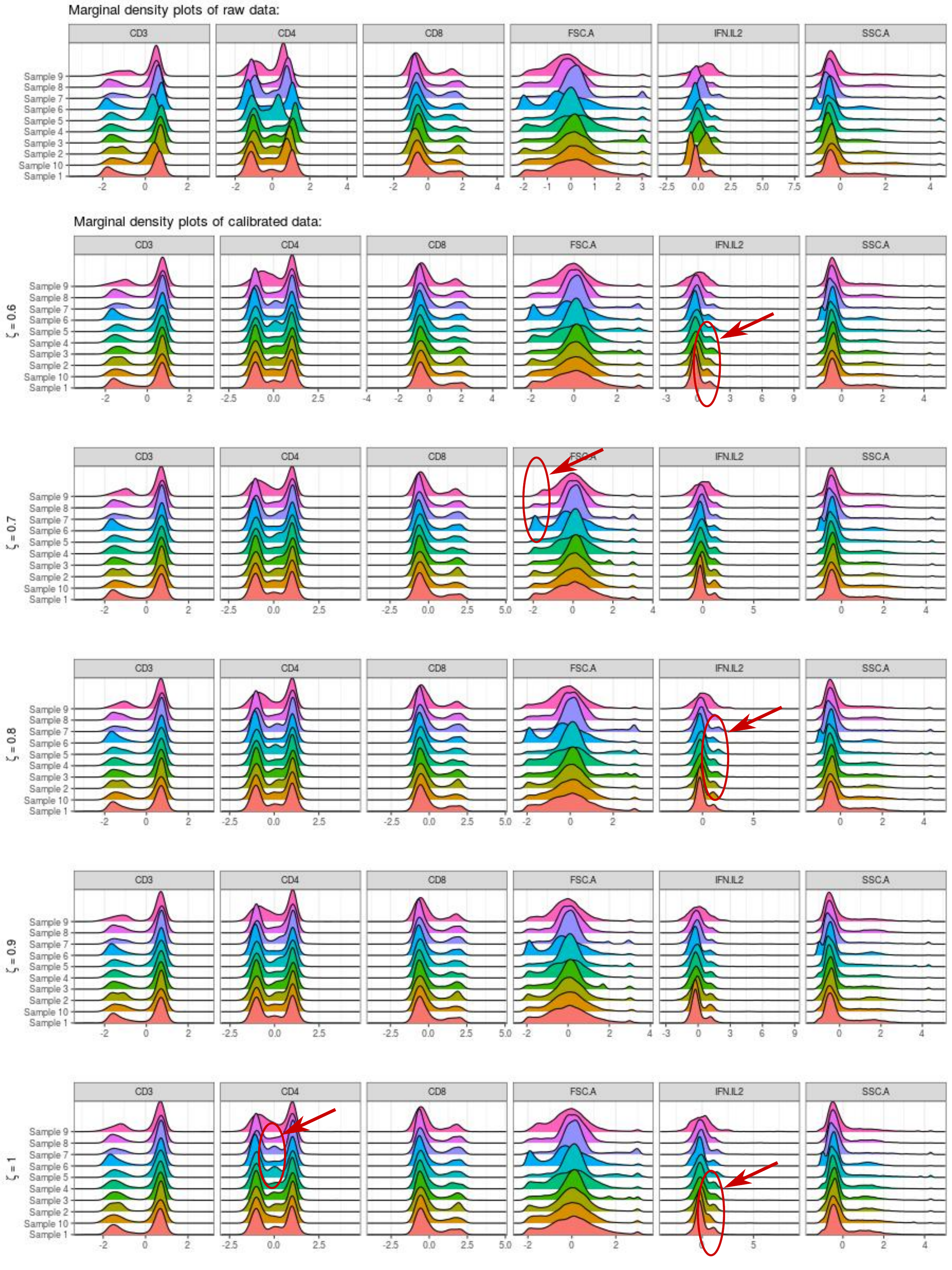}
\caption{Marginal density plots for all markers and samples in the 6-dimensional FCM data set. In red circles we highlight areas in some of the calibrated samples that are not as well aligned compared to the other calibrated samples.\label{fig:sensitivity2}}
\end{figure}

We perform a sensitivity analysis to the values of $\zeta$ for the 6-dimensional EQAPOL data set. We fit the COMIX model for each $\zeta$ in $\{0.1,0.2,0.3,0.4,0.5,0.6,0.7,0.8,0.9,1\}$. In Figures~\ref{fig:sensitivity1} and \ref{fig:sensitivity2} we show marginal density plots for the raw data (top row) and the calibrated data (each row corresponds to a different $\zeta$). Overall, the calibrated samples are much more aligned compared to the raw data. However, for several of the $\zeta$ values we highlighted with red circles some anomalies in the calibration. In addition to the above figures, Table~\ref{tbl:fcm_est_k} presents the number of estimated clusters (number of unique entries in the estimated cluster assignment variable) for each sample and value of $\zeta$. As our data here is a batch control, we may assume that the true number of clusters is the same across all samples. Indeed, the samples for which the estimated number of clusters is the same across all clusters tend to be more aligned and stable compared to the other samples. Based on this sensitivity analysis, values from among $\{0.1, 0.2, 0.5, 0.9\}$ are our best candidates for the value of $\zeta$. The lower values will prefer larger clusters, while the higher one will cater to smaller ones. The choice of $\zeta$ to apply from among those candidates may depend on expert's input as well as the desired emphases for analysis of further data.

\begin{table}[!ht]
\centering
\begin{tabular}{>{\rowmac}c|>{\rowmac}c>{\rowmac}c>{\rowmac}c>{\rowmac}c>{\rowmac}c>{\rowmac}c>{\rowmac}c>{\rowmac}c>{\rowmac}c>{\rowmac}c<{\clearrow}}
$\zeta$ & \#1 & \#2 & \#3 & \#4 & \#5 & \#6 & \#7 & \#8 & \#9 & \#10\\\hline
\setrow{\bfseries}0.1 & 10 & 10 & 10 & 10 & 10 & 10 & 10 & 10 & 10 & 10\\
\setrow{\bfseries}0.2 & 11 & 11 & 11 & 11 & 11 & 11 & 11 & 11 & 11 & 11\\
0.3 & 11 & 11 & 11 & 11 & 11 & 12 & 11 & 11 & 11 & 11\\
0.4 & 12 & 13 & 13 & 12 & 12 & 12 & 13 & 12 & 13 & 11\\
\setrow{\bfseries}0.5 & 13 & 13 & 13 & 13 & 13 & 13 & 13 & 13 & 13 & 13\\
0.6 & 14 & 14 & 14 & 13 & 14 & 14 & 14 & 13 & 13 & 13\\
0.7 & 14 & 14 & 14 & 14 & 14 & 14 & 14 & 14 & 15 & 14\\
0.8 & 15 & 16 & 16 & 15 & 15 & 15 & 16 & 15 & 15 & 15\\
\setrow{\bfseries}0.9 & 13 & 13 & 13 & 13 & 13 & 13 & 13 & 13 & 13 & 13\\
1 & 15 & 15 & 15 & 14 & 14 & 15 & 15 & 14 & 13 & 13
\end{tabular}
\caption{Number of estimated clusters for each sample and value of $\zeta$. Boldface: equal number of estimated clusters across all samples.\label{tbl:fcm_est_k}
}
\end{table}

\end{document}